\documentclass[final,3p,times,twocolumn]{elsarticle}

\usepackage{graphicx}

\usepackage{amssymb}

\biboptions{square, numbers}

\journal{Astroparticle Physics}

\begin{document}

\begin{frontmatter}



\author{Y. Akaike}
 \author{K. Kasahara\corref{kk}}
 \cortext[kk]{Corresponding author}
 \ead{kasahara@icrr.u-tokyo.ac.jp}
\author{ R. Nakamura\fnref{rn}}
 \fntext[rn]{Present address:  Energy and Environmental Systems Lab.,   Energy Optimization Technology Group, NTT, Musashino-shi, Tokyo, Japan}
\author{Y. Shimizu, S. Ozawa and S. Torii}

 \address{RISE,  Waseda University, Shinjuku, Ohkubo, Tokyo, Japan}
  \author{K. Yoshida}
  \address{College of Systems Engineering and Science,   Shibaura Institute of Technology,
  	 Saitama, Japan  }
  \author{T. Tamura and S. Udo}
   \address{Faculty of Engineering, Kanagawa University, Yokohama, Japan}

\title{Feasibility study of probing the high energy end of the primary cosmic electron spectrum\tnoteref{title}
 by detecting 	geo-synchrotron X-rays}
\tnotetext[title]{By electrons, we mean electrons + positrons in this paper.}

\begin{abstract}
Based on tests  of a tentative
detector for observing geo-synchrotron hard X-rays generated by primary electrons,
we study the feasibility  of
probing  cosmic electrons above  a few TeV to over 10 TeV.
  Such high energy electrons are
expected to give  proof of sources near the Earth
(e.g. supernova remnants such as Vela:  age $< 10^5$ years located within $ <$1kpc).
The  idea itself is rather old;   a high energy
electron emits synchrotron  X-rays successively in the geomagnetic
field and thus gives  several X-rays  aligned   on a  meter scale.
This feature is a clue to overcome the background problem  encountered
in other traditional observation methods.  We critically examine  the feasibility 
of this approach  
assuming a
satellite   altitudes  observation,   and  find that it is difficult to derive a precise   energy
spectrum of electrons  but is possible to get a clear signal of the existence of several TeV
electrons   
if the flux  is comparable to the level  predicted by a class of
plausible  models.  
For such observations, an exposure $>1 $m$^2\cdot$year would be needed.  It would be attractive 
to  incorporate the present scheme in  the  gamma-ray  burst observations.

\end{abstract}


\end{frontmatter}


\section{Introduction}
\label{intro}

High energy electrons in space  suffer
 synchrotron and inverse Compton scattering 
 energy loss proportional
to the square of their energy; this means high energy electrons cannot come from
distant sources.
If their energy 
exceeds several TeV, they are
expected to
be accelerated in and come from
supernova remnants (SNR) of age less than $\sim 10^5$ years, located 
within 1 kpc from the Earth.  
Among the small number of known SNR's satisfying   such criteria,
Vela is a good  candidate \cite{velacalc}.  
Detection of such high energy electrons will give  strong support for
particle acceleration in near-by  SNR and will give us important
additional information related to high energy gamma ray observations, 
despite the fact that  the electron  arrival
direction cannot tell the source direction  due to deflection by magnetic fields.

 Observation of high energy electrons  is more difficult than many other primary cosmic rays
 due to the small flux and  large background;  we need large and thick detectors to cope with
small flux and high energy, and special features in detector design to overcome  the huge background
mainly produced by protons.   Therefore, experimental data over $\sim$ TeV is very limited.
Figure 1 summarizea recent observations and some model predictions. 

\begin{figure}[h]
\begin{center}
 \includegraphics[width=7.2cm]{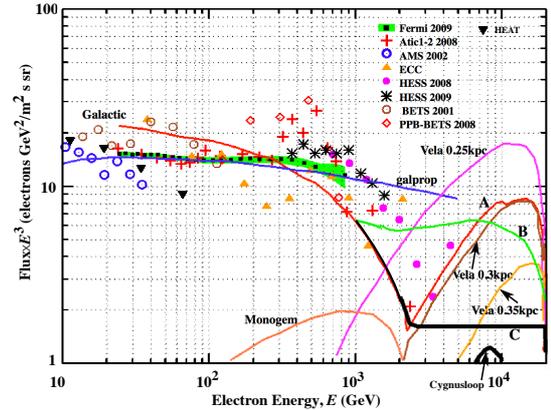}
 \caption{High energy primary electron spectrum obtained by 
Fermi-LAT: \cite{fermi},  Atic1-2:\cite{atic2}, AMS-2001:\cite{ams1},
ECC:\cite{ECC},  HESS-2008:\cite{HESS1}, HESS-2009:\cite{HESS2},
BETS:\cite{bets}, PPB-BETS:\cite{ppbbets}.  Model calculations labeled  
galactic:\cite{galactic} and galprop:\cite{galprop} are for the sum of 
distant contributions.  Others:\cite{velacalc} are possible contributions from
near-by sources (A, B and C are used in the present paper).
}
\label{espec}
 \end{center}
\end{figure}

Recently, the HESS collaboration reported results in the
TeV region by their air Cherenkov light
detection method \cite{HESS1}\cite{HESS2}.  This method is quite different from other conventional direct observation
methods and overcomes the small flux problem.  The HESS  result  indicates a steep 
spectrum, but it is probably premature to conclude the non existence of several TeV electrons\footnote{%
It is interesting to note that their highest energy points are rather consistent with  model A
of the Vela source (Fig.\ref{espec})}.  We will need further investigation.

The CALET project\cite{calet} and AMS-02 project\cite{ams2} are being prepared for observation  on board ISS (International Space Station).   CALET uses a thick calorimeter and will be able to give a  conclusive  result beyond  TeV 
after a few years of  observation.  However, it may turn out that we need to clarify the
10 TeV region flux (for which  CALET may still be  too small) to discuss  near-by sources.
AMS-02 uses a magnet spectrometer and may be  limited for the TeV scale electrons.

The idea of observing geo-synchrotron X-rays  
 emitted successively by  a TeV region electron  and appearing almost on a line, for 
 investigating cosmic primary electrons,   is not new\cite{stephens}\footnote{%
 J. Nishimura had such  an idea around the same time as this reference and recommended us
 to explore the possibility.  Although not easily accessible,
  there are even earlier works \cite{prilutskii}\cite{macbreen}  as mentioned in \cite{stephens}.
 }.
 We study the feasibility of this idea,  paying more attention to the background and 
   stochastic nature of a realistic observation than  earlier works.

The CREST project \cite{crest} which  is based on the same idea  is preparing for a
balloon height observation  around Antarctica.  Its result is highly awaited, and we 
will be able to see consistency with,  or difference from, our  estimation in the present paper, although
we will  discuss a satellite  height  observation.  The CREST observation may not be
 long enough due to the balloon flight limitation;  nevertheless we can expect it will disclose many  
 new features in the geo-synchrotron observation.

\section{Characteristics of geo-synchrotron  X-rays from TeV region electrons}
\begin{figure}[h]
\begin{center}
 \includegraphics[width=6.5cm]{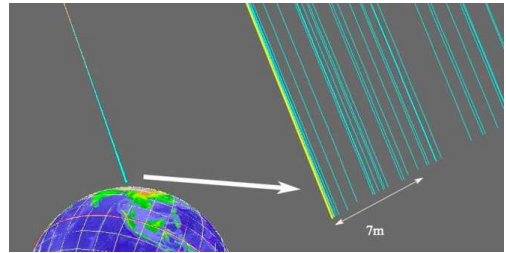}
 \caption{Schematic view of geo-synchrotron X-rays.   Enlarged inlayed figure shows
 an electron (leftmost) and accompanying X-rays with  the density of several/m.
 }
 \label{geosync}
 \end{center}
\end{figure}

Figure \ref{geosync} shows  a schematic view of geo-synchrotron X-rays\footnote{%
We use Cosmos(generation of X-rays)/Epics(detector response)
	 for the present M.C simulation: http://cosmos.n.kanagawa-u.ac.jp. 
The synchrotron X-ray emission is managed by Baring's formula \cite{baring} which
	can be used for our conditions as well as extreme conditions
	 (very high magnetic field and very high energy).
}. 
We briefly recapitulate  the characteristic features of geo-synchrotron X-rays which
are produced by $>$ TeV electrons. 
As is well known, the emissivity of  synchrotron radiation by an electron of energy
$E_e$ shows a peak at $\sim \gamma^2 \nu_c$ where $\gamma=E_e/m_e$ is the gamma
factor of the electron with mass $m_e$, and $\nu_c$ the cyclotron frequency   of the electron
(in $h=c=1$ unit).   At higher X-ray
energies, there is a sharp exponential cutoff in the spectrum. 
The peak emissivity for  the Earth  reaches $\sim$10  keV  for
electrons of energy $\sim$1 TeV.  Therefore, if we observe several X-rays of  energy 10 keV or
more,  we may expect that the 
average energy of
X-rays, $<\!\!E_X\!\!>$, is related to  the  peak energy.   If the average is proportional to
the  peak energy,  $E_e \propto m_e\sqrt{<\!\!E_X\!\!>/\nu_c}$ is expected, but 
this seems to hold only when the minimum X-ray energy is much smaller than the
peak energy.

\begin{figure}[h]
\begin{center}
 \includegraphics[width=7cm]{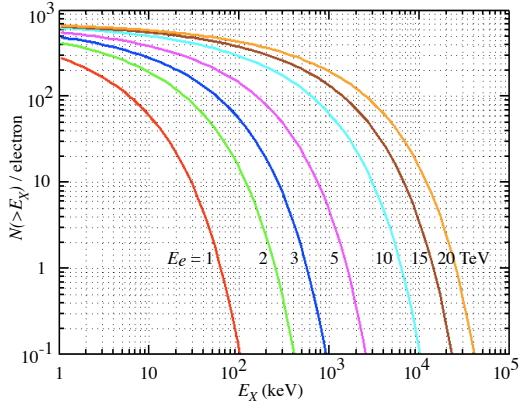}
\caption{Integral energy spectrum of geo-synchrotron X-rays emitted by
1 TeV to 20 TeV electrons}
\label{ExspecI}
\end{center}
\end{figure}
 Figure\ref{ExspecI} shows the average integral 
energy spectrum of X-rays emitted by 1$\sim$20 TeV  electrons which start isotropically from a height of
  $3\times 10^4$ km and are directed to
  height 390 km, 
latitude 0$^\circ$,  longitude 130$^\circ$\footnote{
We used  a geomagnetic filed of year  2010 by referring to
the year 2005 IGRF data; http://www.ngdc.noaa.gov/IAGA/vmod/igrf.html
}.

\begin{figure}[h]
\begin{center}
 \includegraphics[width=6.5cm]{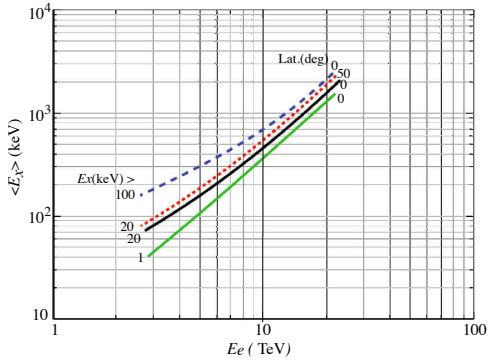}
 \caption{Average energy of X-rays  vs electron energy.
 The minimum X-ray energy is 1, 20, 100 keV as indicated
 in the Fig.   For minimum energy of  20 keV, results for
 latitude 0$^\circ$ and 50$^\circ$  are shown.
 }
 \label{EeVSaveEx}
 \end{center}
 \end{figure}

The number of X-rays and the spectrum shape depends on the location.  At higher latitudes,
in general, 
a harder spectrum is seen, and the  total number of X-rays becomes smaller.  However,
such  differences can be neglected  for our present feasibility study, and
we  use this observation point for further discussions  in this article\footnote{
The background depends much more strongly on the latitude so we consider it
separately.}.

In Fig. \ref{EeVSaveEx}, the average X-ray energy is plotted as
a function of the electron energy.  The minimum X-ray energy  is
varied from 1 keV to 100 keV to see the change of the energy dependence.
The result for latitude 50$^\circ$ is also shown to illustrate the small 
difference from the latitude 0$^\circ$ case.  For 10 TeV electrons,
the average X-ray energy is 600$\sim$700 keV. If the  number of
X-rays is limited to several,  the average generally becomes smaller.

\begin{figure}[h]
\begin{center}
 \includegraphics[width=6.5cm]{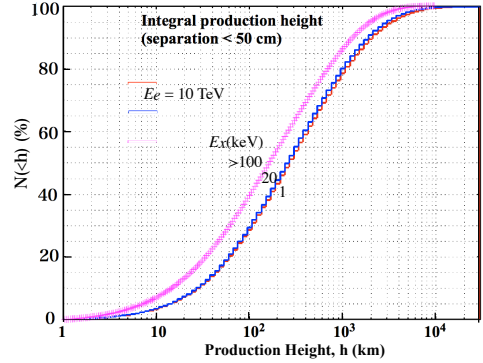}
\caption{Integral production height distribution of those X-rays 
that have another X-ray within 50 cm.
 }
\label{ProductionH}
\end{center}
\end{figure}

 The integral production height of X-rays is shown in Fig. \ref{ProductionH}.  
We chose those X-rays that have another X-ray within 50 cm. The
production height less than 500 km contains 60 \% of the events.  
To judge that X-rays observed are due to the geo-synchrotron emissions,   we impose
the condition
that several X-rays come simultaneously on a line.  Since the emission angle
of the synchrotron X-rays is order of  $\sim 1/\gamma$,  the deviation  from 
a line alignment  is $\sim$ 2.5 cm for X-rays coming from electrons of energy  $E_e=10$ TeV  at height 
500 km. 
  If we impose several
X-rays to be aligned in a small distance,   lower altitudes with
higher magnetic field strength become  more important.   Therefore, 
5 cm is a good measure for the  position resolution of  the  X-ray detector. 

\begin{figure}[h]
\begin{center}
 \includegraphics[width=7cm]{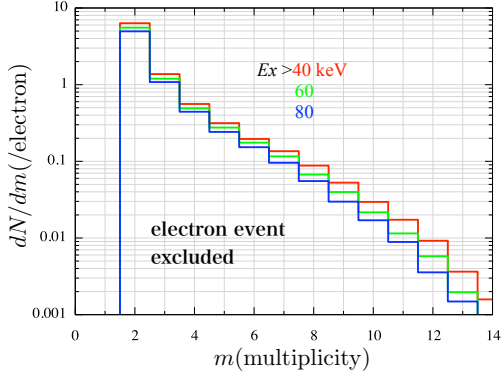}
 \caption{Distribution of  X-ray multiplicity  in  1.4 m $\times$ 1.4 m area
 for a primary electron  of 10 TeV}
\label{m-dist}
 \end{center}
\end{figure}

In Fig. \ref{m-dist}, we show the multiplicity distribution of X-rays falling in an
area of 1.4 m$\times 1.4$ m for a primary electron of energy 10 TeV. The
minimum energy of X-rays is set to be 40, 60 and 80 keV.
 In the distribution, we exclude the case where
the electron itself enters the area; this is because the electron emits
a number of photons by radiation in the detector  and   they are observed
simultaneously at various positions together with the X-rays.
This makes it difficult to judge  the intrinsic 
X-rays as  discussed later (section \ref{sec:cp}, Fig. \ref{splash})

\section{Consideration of the detector efficiency and backgrounds}
For further analysis, we have to assume  a model detector and stochastic nature of the observation,  together
with a  background  for the observation.  Our assumptions are based on a basic test experiment 
summarized in the Appendix.  For example, we put an error to the energy, $E$ (keV), absorbed in BGO  
by a random sampling with FWHM error of 16$\sqrt{662\ {\rm keV}/E}$ \%.

\subsection{The tentative detector}
\label{sec:detector}
The tentative detector is summarized as follows:
\begin{enumerate}
\item The unit X-ray detector is made of BGO. The cross-section is 5 cm $\times$ 5 cm  and the
	thickness  2 cm (1.79 r.l;  1r.l = 1.12 cm).
	   Other BGO properties are: density 7.13 g/cm$^3$,  
        light yield 8.2 photons/keV or $15\sim 20$ \% of NaI,
       decay time 300 ns.
\item Each BGO is wrapped in teflon. The top part is covered by CFRP (Carbon Fibre Reinforced Plastic) 
	of 500 $\mu$m thick.
\item The gap between  BGO  units  is filled with thin (1 mm) Sn. Otherwise, 
    a	high energy albedo  gamma ray could run the  small gap rather a long distance 
  and cause multi-Compton scattering  which could be a major background. 

\item   The whole  detector size is 1.43 m $\times$ 1.43 m  (28 $\times$ 28 BGO's)
\item  Each BGO signal  is read by a PMT at the bottom.  
\item  Simple supporting  platform is assumed together with material
	representing PMT (Fe and glass).
\end{enumerate}

\subsection{The observation conditions}
\label{sec:obcon}
We impose the following cuts on each event.
\begin{enumerate}
\item  We put a threshold,  $E_{th}$  on the observed energy, $E_X$, in each unit BGO. The 
standard value of $E_{th}$ is 80 keV.
   Those BGO's with observed energy $E_X<E_{th}$ 
	are neglected.
\item 	The number of such BGO units (i.e, $E_X>E_{th}$), $m$,   must be $\ge 5$.
            (Of course, we cannot know multiple incidence of X-rays in one BGO and 
            	such X-rays  are  regarded as a single X-ray, though the probability is very
		small). 
	   
	\item   The energy  sum of the two highest energy BGO's ($E_1+E_2)$ must be $<4.5$ MeV.  This is to
	suppress   background events due to multi-Compton  scattering  of  high energy
	gamma rays. 
\item  \label{sec:BGOblock}
  Let the maximum distance among  $m-$BGO's be $d$ cm.   The center of each BGO
is used to calculate the distance.  $d$ is converted to  an integer $b$ expressing
 an effective  number of  BGO blocks by $(5.1b+5)=d$.  We require $b\ge 11$ 
 (See section \ref{sec:SN}).
  Hereafter $b6$ means $b=6$.

\end{enumerate}

\subsection{Collection power}
\label{sec:cp}
\begin{figure}[h]
\begin{center}
 \includegraphics[width=7cm]{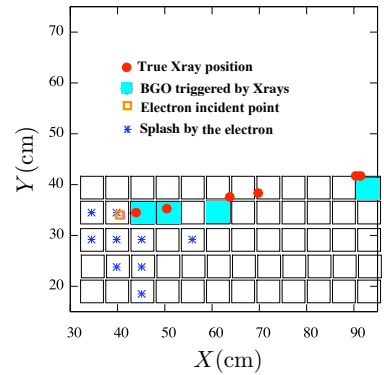}
 \caption{The electron splashes a lot of 
 photons to trigger a number of BGO's
 }\label{splash}
 \end{center}
\end{figure}

One of the merits of observing geo-synchrotron X-rays has been believed to be 
  that we don't need to observe
electrons directly and hence  the effective area could be larger than that for direct observation.
We define the collection power, $C_p$, by
\begin{equation}
C_p =\frac{ {\rm Number\ of\  events\ with\  a\ desired\ condition } }
{{\rm  Those\ events \ containing\  the\ electron }}
\end{equation}
We note that we must exclude  an event  which contains 
the electron itself, since the high energy electron ($>$ TeV) splashes  a number of 
photons  to surrounding BGO's and
it becomes difficult to identify synchrotron  X-rays  (Fig. \ref{splash}).  

This fact  seems to have been overlooked in   past papers,  and 
   unfortunately, reduces $C_p$ a large amount.   If we use a loose cut for 
   observation, $C_p$ could be as large as 5, but  our condition 
    reduces $C_p$ to $<1$.  This is partly due to  the fact that
    we require $m\ge 5$; such high multiplicity X-rays  tend to be
   generated near the Earth where  the magnetic field is strong, and
   hence separation of the X-rays and the electron is not large; that means
   many of the  high multiplicity events are accompanied by the electron.

For our observation conditions,  events containing the electron are  automatically 
excluded since the  energy loss in BGO's exceeds 18 MeV and
$E_1+E_2> 4.5 $MeV.
In the present paper, we don't pursue how to utilize such events; if we could find
a  method for utilizing the events, $C_p$ will increase substantially.

 \subsection{Dependence of observables on  electron energy }

\begin{figure}[t]
\begin{center}
 \includegraphics[width=7cm]{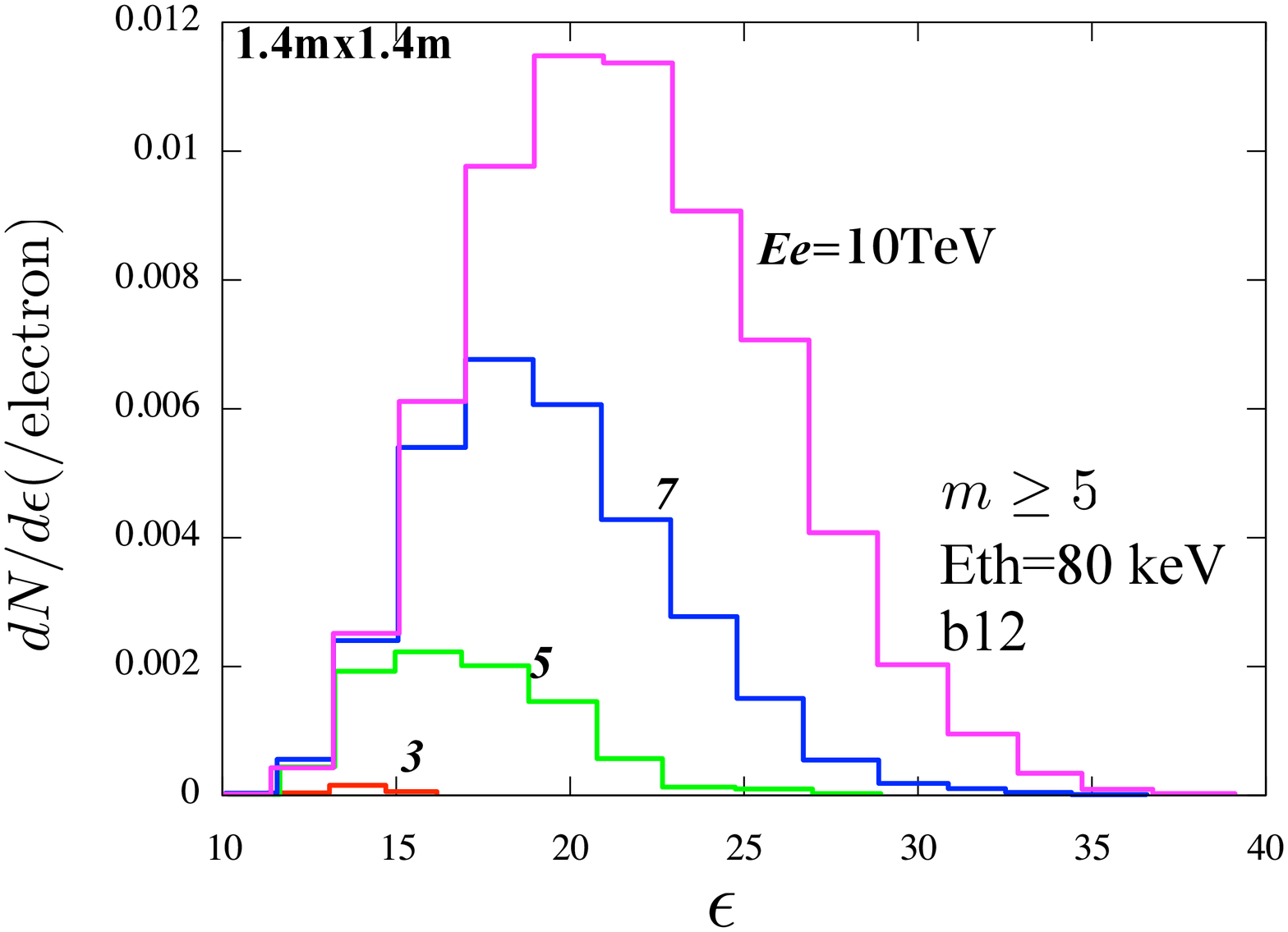} \includegraphics[width=6.75cm]{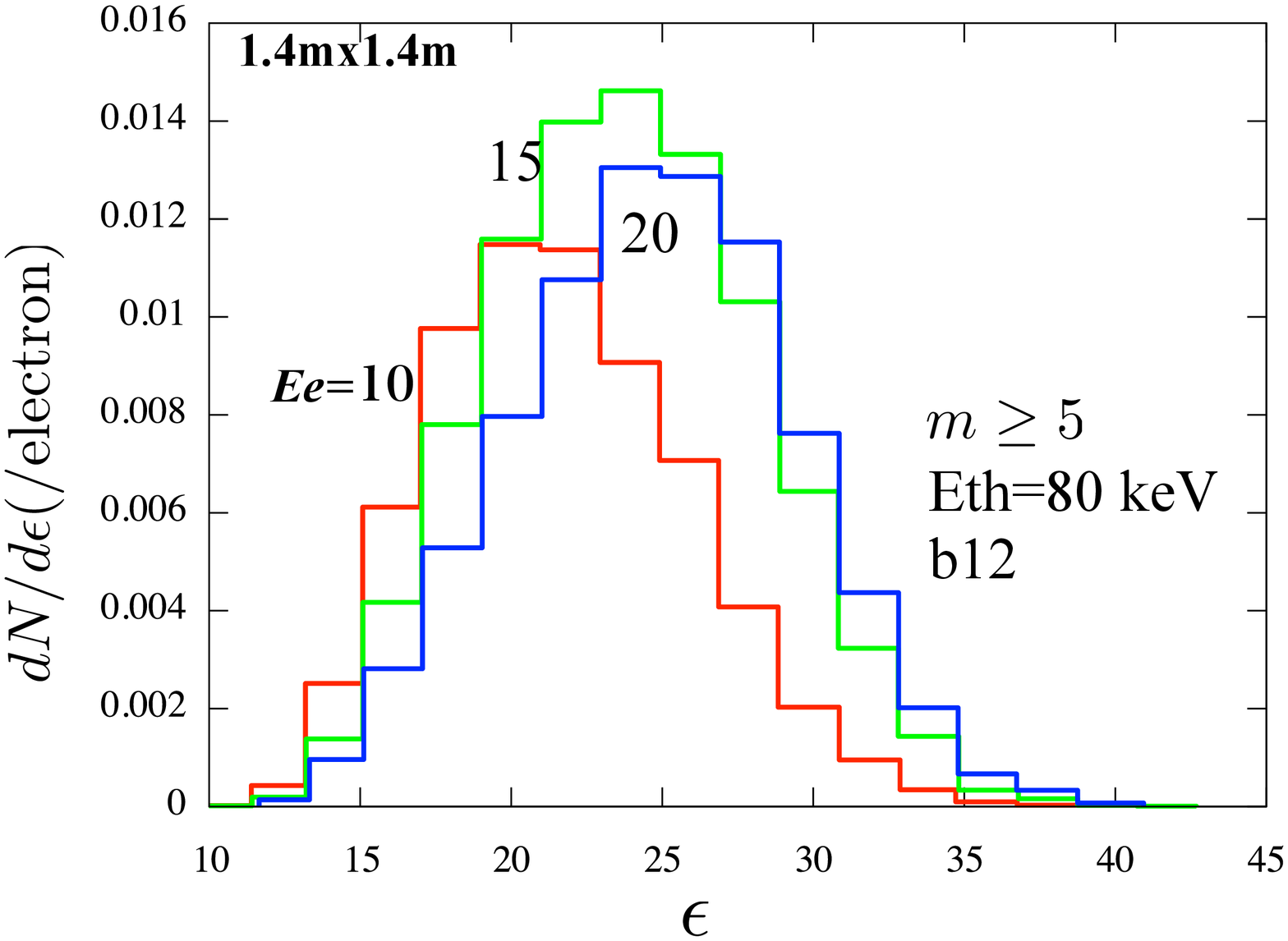}
 \caption{$\epsilon$ distribution satisfying the observation conditions.  Upper: for electron
 energy,$E_e=$ 3, 5, 7 and 10 TeV.  Lower:  $E_e=10,15$ and 20 TeV.  }
\label{epsVSEe}
 \end{center}
\end{figure}

We  use the
average of square root  of  observed  energy as a representative 
of $m-$BGO triggered by X-rays\footnote{If we use a loose  cut,  the square root of X-ray energy  is roughly
proportional to the electron energy. However, this merit is lost  by various cuts imposed 
on events.   Using the simple average or median value of energy would lead to the same
conclusion as the present one.}.   The energy is in keV unit.
\begin{equation}
 \epsilon = <\!\sqrt{E_X/\rm{keV}}\! >
\end{equation}
The distribution of $\epsilon$ for various electron energies are
shown in Fig. \ref{epsVSEe} for $m\ge 5$, $E_{th}=80$keV, and $b12$ (see \ref{sec:BGOblock} in sec.\ref{sec:obcon}) .

The peak position, $\epsilon_p$,  of  each $\epsilon$ distribution 
and the electron energy are connected roughly by
$E_e\sim (\epsilon_p/10)^3$ TeV.  The width of the distribution 
is large ($\sim$ 80 \% in FWHM)  so that it would be difficult to
derive a precise energy spectrum of electrons from those observations.

However,  there is a sharp exponential-like cutoff at the high energy side of the synchrotron
 X-ray spectrum and this gives us, in a sense,  the function of a threshold  counter to our
 detector.  Actually, we see there is no sensitivity to electrons with 
 energy lower than 2 TeV for the standard cuts.  
In  Fig. \ref{epsVSEe}, each distribution is normalized to give $C_p$   
when it is  integrated. The values are 
 summarized in Table \ref{epstab}.

\setlength{\tabcolsep}{1.5mm} 
  \begin{table}[htdp]
\caption{Dependence of $C_p$ on electron energy}
\begin{center}
\begin{tabular}{|c|c|c|c|c|c|c|}
\hline
$E_e$(TeV) &　3 & 5 & 7 & 10 & 15 & 20  \\
\hline
$C_p$ & 0.0005  & 0.018 & 0.06 & 0.13 &  0.18 & 0.18  \\
 \hline
\end{tabular}
\end{center}
\label{epstab}
\end{table}%

\setlength{\tabcolsep}{2mm} 

\subsection{Case study of plausible models}
We use 3 models, A, B \cite{velacalc} and C, for the electron energy spectrum in the
highest energy region  as shown
in Fig. \ref{espec}. 

Vela is a good candidate for  high energy electron sources.  Some models
predict that $\sim$ 10 TeV electrons are reaching the Earth.  
The prediction depends on the distance to Vela,  diffusion constant etc.
For example, distance dependence is shown in Fig. \ref{espec} for 0.25, 0.30  and 0.35 kpc
and a diffusion constant of $D=1.0\times 10^{29}(E/$TeV)$^{0.3}$cm$^2/$s. 
An  estimate of the  probable  distance  to Vela is 0.25 kpc  while the maximum value 
is expected within  $0.39\pm 0.1$ kpc \cite{veladist}.  Since other factors also
affect the flux from Vela, we employ  here the 
rather conservative (with respect to flux) distance of  0.3 kpc and use
the flux including the contribution from Cygnus Loop.  We call this model A.
Larger diffusion constants flatten the spectrum.  As an example of such a case, we 
use model B as shown in Fig. \ref{espec}.    Model C is a completely artificial spectrum
to examine the detection limit.
 
 The line labeled 'galactic' is a model calculation \cite{galactic} which includes distant sources as 
 well as possible  near-by  sources such as Monogem.  The line labeled 'galprop'  is another
 such model \cite{galprop} calculated by the Galprop program.  
 
 Table \ref{velaflux} shows the integral flux for model A, B and C. 
 \begin{table}[htdp]
\caption{Integral flux of model A, B and C:  electrons/$\pi$ sr$\cdot$m$^2\cdot$year}
\begin{center}
\begin{tabular}{|c|c|c|c|c|c|c|}
\hline
 & \multicolumn{6}{|c|}{Energy(TeV)}\\
\cline{2-7}
model & $>1$ & $>3$ & $>5$  & $ >7$ &  $>10$ & $>15$ \\
\hline
 A &  2248 &    243 &  129 &  74 &   32.8 &    8.1 \\
B &   3021  &      324  &   112 &    50 &    18 &     3.6 \\
 C &  2080  &    86  &    30  &   14 &     6  &      1.6 \\
 \hline
\end{tabular}
\end{center}
\label{velaflux}
\end{table}

\subsection{Background considerations}
When we take  an event with $m(\ge$5) synchrotron X-rays  aligned,
possible backgrounds could come from
\begin{enumerate}
\item Chance coincidence of X-rays or charged particles 
\item Multiple Compton scattering  by a high energy  X- or gamma-ray entering   the detector 
\end{enumerate}
Possible  sources of such  backgrounds would  be
\begin{enumerate}
\item Uniform isotropic X-rays and gamma-rays ( Cosmic  X-ray Background or CXB)
\item Cosmic ray charged  particles such as electrons and protons
\item X-rays from the galactic  plane and strong point sources
\item Albedo particles from the earth's atmosphere
\item  A Gamma ray burst
\item   Radiation from the environment surrounding   the detector
\item  South Atlantic Anomaly Radiation
\item A  Solar flare
\end{enumerate}
As to the environmental  radiation,  we must choose a low background environment
and, this is expected to be possible owing to our observation conditions.
The last two sources might be  avoided by switching off the electronics. If a gamma ray burst
is as strong as triggering more than 5 BGO's  almost on a line in a short time
(10 ns; see later), we could get information from other GRB observatories.

Next, we summarize other factors  and derive the chance coincidence rate.

\subsubsection{Cosmic X-ray Background: CXB}
CXB (Diffuse X-rays) is now well understood and we can calculate the background   due to CXB.   The background
by other sources is estimated as the ratio to the one for CXB. 
As the low energy CXB, we use the data from HEAO-I \cite{heao} 
which was  confirmed
by Swift/BAT \cite{swift} recently. The data can be continued to the
SAS-2 \cite{sas-2b} and EGRET \cite{egret2b}\cite{egret2c} data at higher energies (Fig. \ref{fig:cbxVSalbedo})\footnote{
Although energy is too high for our interest, 
a recent Fermi-LAT \cite{fermilat} result over several 100 MeV region gives a lower intensity than
\cite{egret2c}  which gives little bit lower revised flux than the original one \cite{egret2b} at several 10 MeV.
}.

Chance coincidence by CXB must be considered at low energies ($< 500$ keV). As to 
low  energy electrons and protons,  their flux \cite{nasa1}\cite{nasa2}
is
 much lower than CXB
 and we don't need to consider
their effect.  Also it's difficult for high energy charged particles to
generate BGO signals on a line with energy deposit less than a few MeV which is our
maximum cut.

 \begin{table*}[htdp]
\caption{Flux of CXB: the 3rd line shows the integral flux over $E_X$(in $\pi$sr)}
\begin{center}
\begin{tabular}{|c|c|c|c|c|c|c|c|c|c|c|}
\hline
$E_X$ (MeV) & 0.01  &   0.02 & 0.03 &  0.04 & 0.05 &  0.06 & 0.07 & 0.08 &0.1 &   0.2     \\
\hline
/cm$^2$s sr MeV  &  316   &  102 &  42.2 & 25.6 & 15.0 & 9.3 & 6.15 & 4.33 & 2.43 &  0.417 \\
\hline
/cm$^2$s   & 10.7 &  5.18 &  3.12 &  2.09 & 1.47 &  1.10 & 0.86 & 0.704 & 0.5 & 0.175  \\ 
\hline
\end{tabular}
\end{center}
\label{diffuseX}

 \begin{center}
 \begin{tabular}{|c|c|c|c|c|c|c|c|c|}
 \hline
   0.3 &  0.5 &           1   &          2  &             5 &    10  &      20  & 50 &100\\
\hline
0.15 &  0.042  & 7.4(-3)  & 1.4(-3) & 1.5(-4) & 2.8(-5) & 5.9(-6) & 7.9(-7) & 1.6(-7) \\
 \hline
0.096 & 0.045  &  1.69(-2)  & 6.6(-3)   & 1.75(-3)  & 7.2(-4)  & 3.08(-4) & 1.01(-4)  &  4.5(-5)\\
 \hline
\end{tabular}\\
\flushleft{\vspace{-0.15cm}\hspace{1.cm}(-2) etc means 10$^{-2}$ etc．}
 \end{center}
\end{table*}

\subsubsection{Atmospheric albedo particles}
High flux low energy particles which might contribute to chance coincidences 
can be absorbed by the support structure to be put on the Earth side. High energy particles
are not able to make  fake events by multi-Compton scattering.  However, X-rays and 
gamma rays in the intermediate energy region
must be   considered.  Their angular distribution and latitude 
dependence  are rather complex. At high  energies ($> 30$ MeV), there are
data from  SAS-2 \cite{sas-2} and EGRET \cite{egret}.   In the X-ray energy region, recent reliable observations from
Swift/BAT \cite{swift} show a steeper spectrum and thus a higher flux
in the low energy region ($<200$ keV) than the older one \cite{imhof}.   

\begin{figure}[h]
\begin{center}
\includegraphics[width=7.5cm]{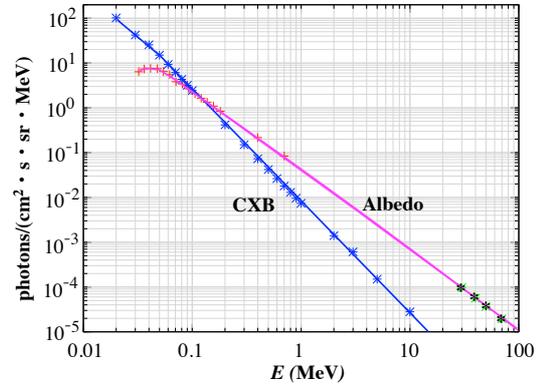} 
\caption{Energy spectra of the isotropic diffuse X/gamma-rays (CXB) and atmospheric albedo X/gamma-rays.}
\label{fig:cbxVSalbedo}
\end{center}
\end{figure}

Although there is  a maximum of  5 times  difference  in the  fluxes  
around the equator  and magnetic poles, the over all difference is
within a factor of 3. 
We smoothly connect Swift/BAT data  averaged over latitude and SAS-2($\sim$ 30 MeV) and
 EGRET ($\sim$ 100 MeV) data. 
  Figure \ref{fig:cbxVSalbedo} shows the energy spectra
 of CXB and albedo X/gamma-rays which we used in the simulation. 
  At the high energy side, the albedo flux 
 is much higher than CXB.  Since the atmospheric  albedo is complex, 
 we   use, for safety,   an  overestimated flux 
  if there are uncertain factors. 
 
 From this point of view, 
 we employ a 1.5 times higher  flux than 
Swift/BAT data in Fig. \ref{fig:cbxVSalbedo}.
 We also assume that the Swift/BAT data is the flux from the
 Nadir direction (this will probably  lead to overestimation of the background). 
  If we measure the X/gamma-ray  angle $\theta$ from the  Nadir,  the   Earth horizon
 is at $\theta \sim 70^\circ$ where the albedo flux becomes maximum.  This enhancement effect
 can be well modelled by assuming that the angular distribution of  X/gamma-rays which enter the
 bottom side of the detector is not $\cos\theta d\cos\theta$ but $d\cos\theta$ (i.e, not 
 isotropic but there is  $\sim\sec\theta$ enhancement).

With this assumption, 	if we compare the all angle intensity of the albedo over 80 keV to CXB
	without considering  absorption by the structural material,  we get
	albedo/CXB  $\sim 4$.  If we assume a supporting material 
	equivalent to Pb 500 $\mu$m or Fe 1 mm, the albedo 
	flux becomes 1.29 (/cm$^2\cdot$s) or 2.15 (/cm$^2\cdot$s), respectively.
	The ratio to CXB reduces to $\sim 1.8$  or $\sim 3.1$, respectively.
	However, we don't consider this absorption effect for albedo for the moment. 
	
\subsubsection{X-rays from the galactic plane}
The X-ray intensity per unit solid angle from the  galactic plane is factor  $\sim$10
stronger than CXB.  However, what is important for the chance coincidence 
is the integral value over a hemisphere. Its ratio to CXB is considered to
be almost constant above 10 keV to over 10 MeV. This ratio  is 
estimated to be less than 0.1 by Ginga\cite{ginga} and
ASCA\cite{asca}.   This value is for the case when
the galactic plane  is near the zenith.  
Therefore, we can safely neglect  X-ray contribution
from the galactic plane as compared to 
albedo or CXB.

\subsubsection{Chance coincidence}
\begin{figure}[h]
\begin{center}
\includegraphics[width=7cm]{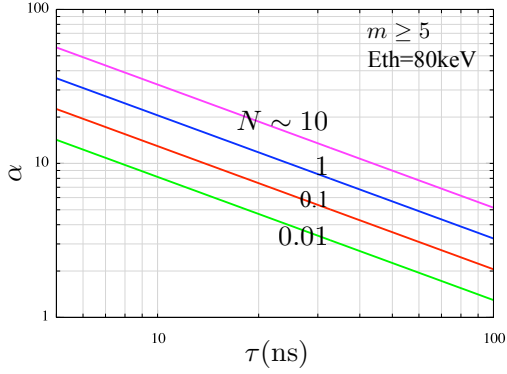} 
\caption{The number of  chance coincidence  events, $N$,  as a function of time resolution ($\tau$)
and effective background index($\alpha$) for 
$S$=2m$^2$, $T=1$ year．
}
\label{fig:chance}
\end{center}

\end{figure}

We first consider the chance coincidence by CXB;
we estimate  background  events  with $m$ X-rays
aligned.   The factors to be considered are
\begin{itemize}
\item  Time resolution of the system.  $\tau$ in FWHM.
\item  Effective flux over 80 keV.  $f$
\item  Area where the X-rays there can be regarded as on a line.  $\Delta S$
\item  The total area of the detector. $S$.
\item  Observation time.  $T$.
\item   X-ray multiplicity.  $m$.
\end{itemize}
Then, the number of chance events, $N$, in the observation time is
\begin{equation}
N \sim fST(f\Delta S2\tau)^{m-1} 
\end{equation}
Expressing  $\tau$  in ns, 
	 area in m$^2$, $T$ in year
and  $f=\alpha f_0$ where $f_0=0.704$/cm$^2\cdot$s is the CXB integral flux over
 80 keV (we call $\alpha$  effective background index),
	 we obtain
	\begin{equation}
N \sim 2.18\alpha ST(\alpha\Delta S\tau)^{m-1}(1.408\times 10^{-5})^{m-1}\times 10^{11}
\end{equation}
For $S=2$, $\Delta S=0.2$, $T=1$, $m=5$, 
\begin{equation}
N \sim 2.74\times 10^{-11}\alpha^5 \tau^4
\end{equation}
This relation is shown in Fig. \ref{fig:chance}

As described later, we found $\tau=5.4$ ns with use of a constant fraction discriminator; 
this value is realistic even at 80 keV but we assume here $\tau=$10 ns for  safety.
The maximum possible  value of $\alpha$ would be 

 \[  \alpha =1(\rm{CXB}) + 4(\rm{albedo}) + 0.1(\rm{galactic\  plane})\sim 5  \]
 
 Then, we can expect the number of  chance coincidence events is  $\ll 0.01$ for a 1 year observation.
As the condition for line alignment, we use only the $\Delta S$ constraint, so the actual
background could be reduced.

 \subsubsection{Background by multi-Compton scattering}
 
A gamma-ray   may repeat a Compton scattering in BGO and/or environmental
 media and can make a fake event as if synchrotron  X-rays are  on a line. 
Due to the condition, $m\ge 5$,
the energy of such a gamma-ray is concentrated  in $5\pm 3$ MeV
and very seldom exceeds 25 MeV.

If we require  $m\ge 5$ and the maximum distance between the
triggered BGO's be greater than 30 cm, we may expect a very small number of
fake events.
However, the number of CXB over 1 MeV entering
2 m$^2$ area in one year exceeds  $10^{10}$ while 
expected  signal is order of  $10\sim 20$ events.  Then,
we have to consider an event with probability of  $\sim 1/10^9$.

We first calculate multi-Compton event rate using CXB, and 
the contribution from albedo is considered by  introducing the effective 
ratio $\displaystyle\alpha'=\frac{albedo}{CXB} $ 

Multi-Compton events are efficiently  produced by $\sim$ 5 MeV gamma-rays.
The albedo gamma-rays in this energy range are $\sim 40$ times more
abundant than CXB. 
Albedo from   the Earth horizon have a nadir angle of 
$\theta\sim 70$ degrees and  are apt to  produce multi-Compton scattering events.
Therefor, it would be appropriate  to take   $\alpha'\sim 40\times 3\pi/4 \sim 100$ for
safety\footnote{%
$\sin\theta$ is a good enhancement factor for multi-Compton scattering ($\theta$ is
a nadir or zenith angle).  Then, albedo enhancement factor  relative to CXB is 
$\int \sin\theta d\cos\theta/\int \sin\theta \cos\theta d\cos\theta = 3\pi/4$
where the integration region from 0 to $2\pi$ may be used.
}.

\begin{figure}[htbp]
\begin{center}
　\includegraphics[width=68mm]{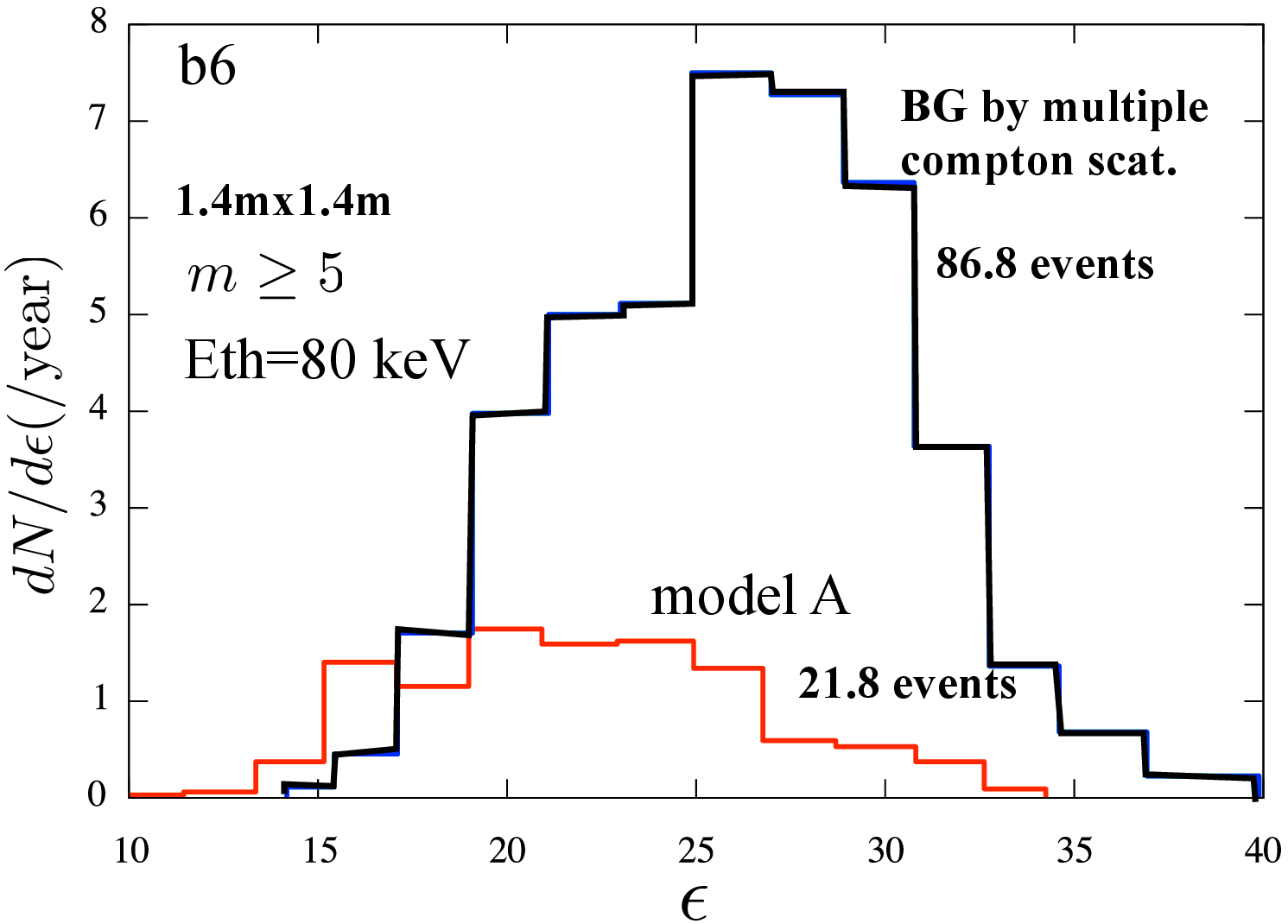}
　\includegraphics[width=68mm]{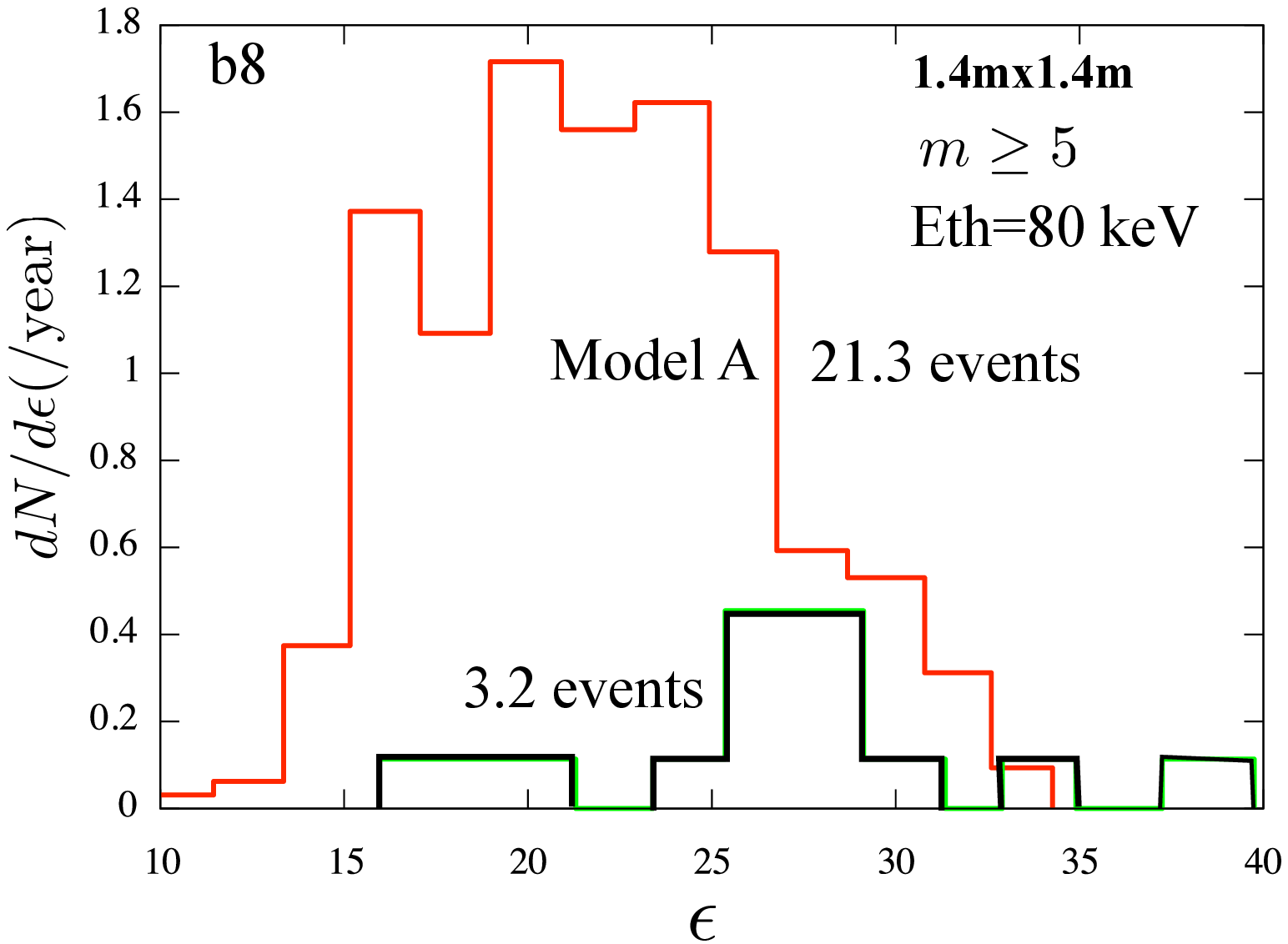}\\
　\includegraphics[width=68mm]{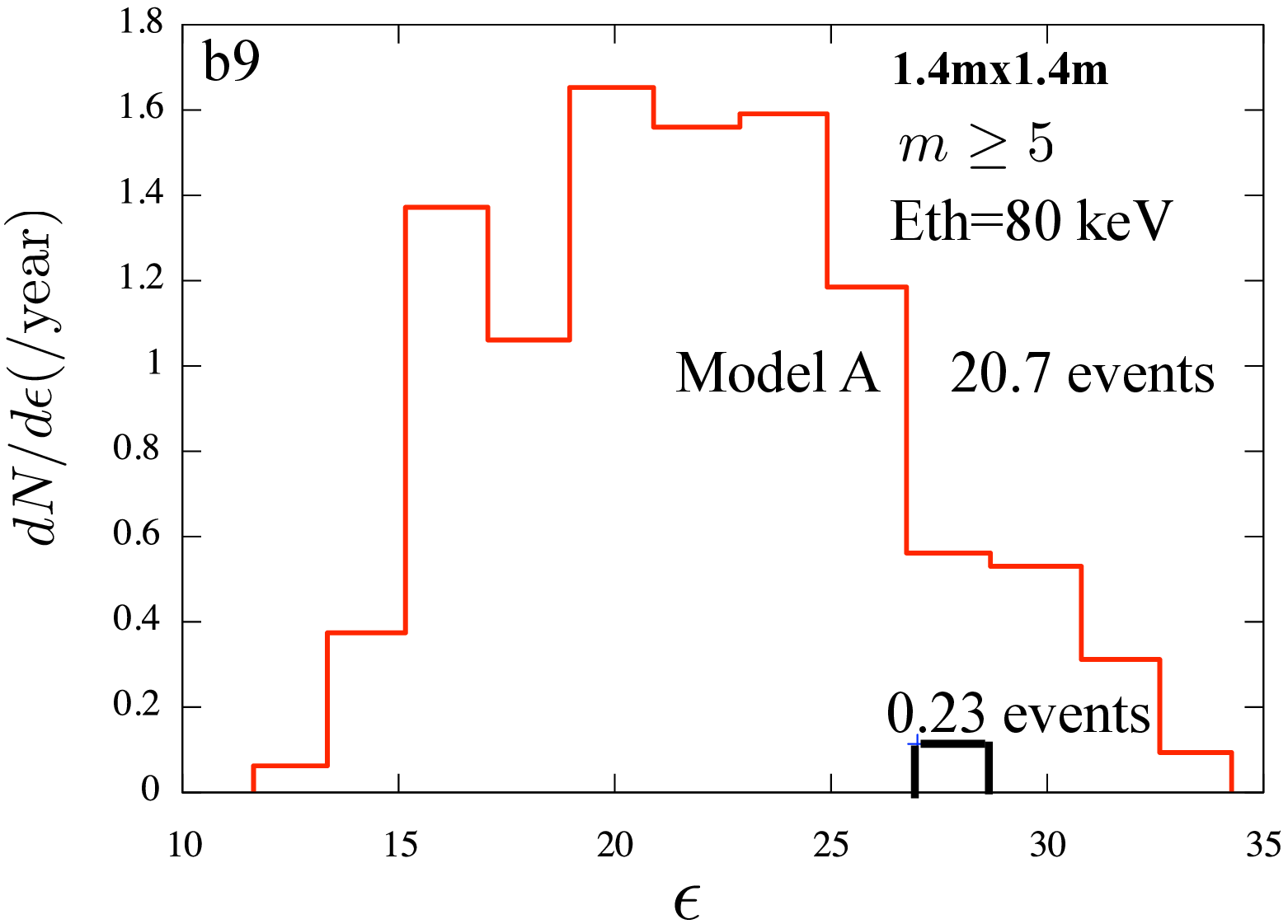}
　\includegraphics[width=68mm]{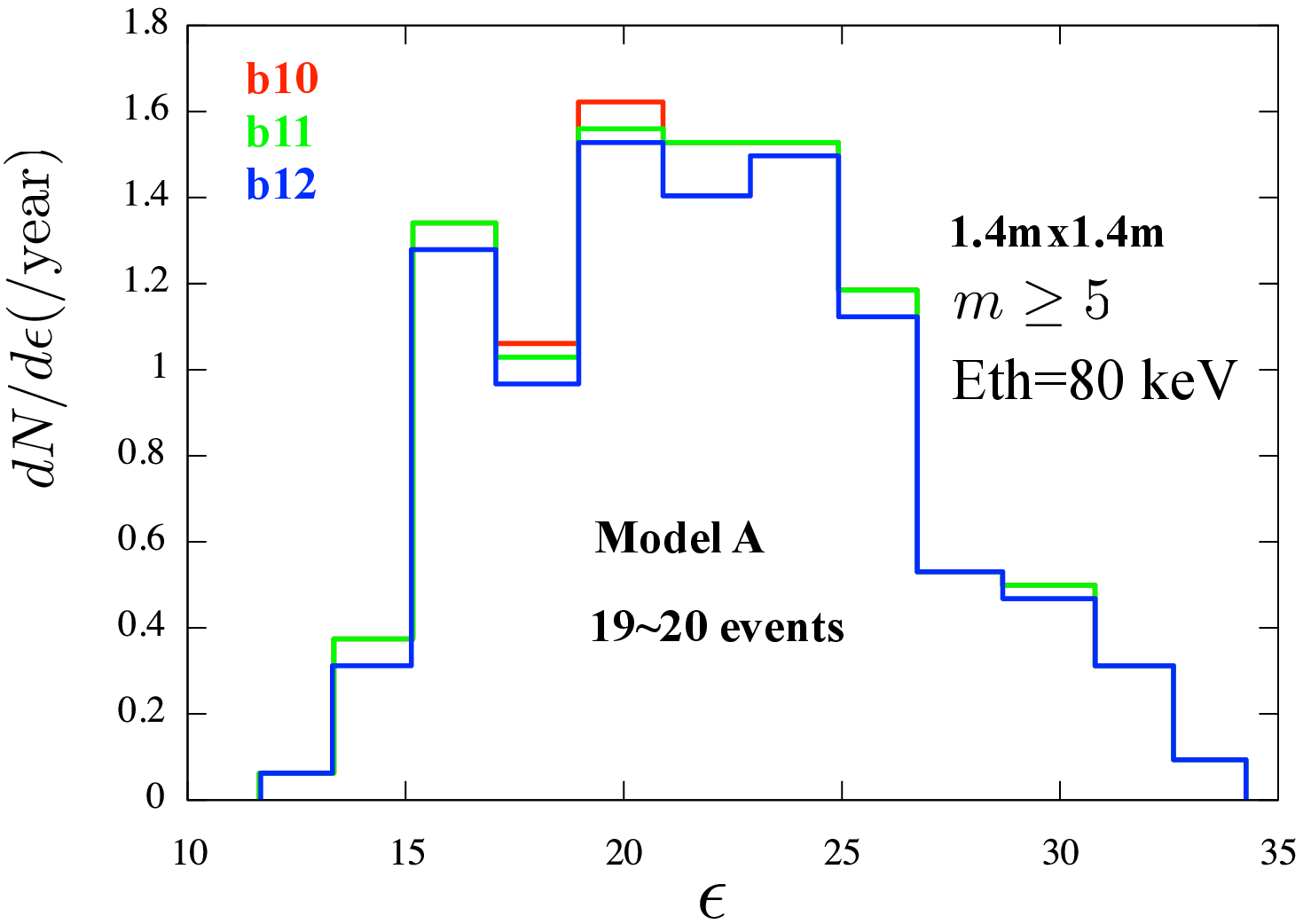}

  \caption{ $\epsilon$ distribution by model A and background for
  	 one year observation. The background rapidly reduces as
	 $b$ increases.  
	  Top: $b6$, background dominates. 
	  next: $b8$,$b9$,  
     	   bottom: $b10,11,12$.  background vanishes but 
	 	signal remains almost the same.
	}
\label{modelAvsBG}
\end{center}
\end{figure}

\subsubsection{Signal vs background}
\label{sec:SN}
To show how the aligned X-rays are widely spanned, 
we use, for example,  $b10$ as defined in  \ref{sec:BGOblock} in section \ref{sec:obcon}.

Figure \ref{modelAvsBG} shows signal vs background for model A under
the condition of
$S=2$m$^2$, $E_{th}=$80 keV, $m \ge 5$ for various $b$\footnote{%
Simulations corresponding to several years are converted to
one year  equivalent.  In this simulation, only CXB with energy greater than
1 MeV is considered.}. 
The CXB background for $b9$ becomes $\sim 0.2$ but
if the albedo effect is to be included, $\alpha'$ must be
multiplied and the background becomes quite comparable to
the signal ($\sim 20$ events).
The signals for b10 to b12  do not change 
appreciably, while background decreases faster than
exponential  and for b10,11,12 it becomes $\sim$0.02, 0.001,
 $\ll\!0.001$\footnote{%
This faster-than-exponential-decrease is realized only if  we fill
space between  BGO's by, say, Sn.  If there is 2 mm unfilled space
between BGO's,  the event rate (/year) becomes $\sim 300\exp(-1.15(b-5))$
and even b12 will receive  albedo  background of order $\sim 10$.  
With the filled space,  background rate for $b>6$ is expressed by
$\sim 300\exp(-1.15(x-5)^{1.30})$.  This relation is inferred by extrapolating
the calculations done up to  b10 (calculations for b11 and b12 need huge
compter power).
}.
Therefore, with b11 or  b12, we can completely neglect the background
from  albedo multi-Compton scattering. 

\begin{figure}[htbp]
\begin{center}
\includegraphics[width=7cm]{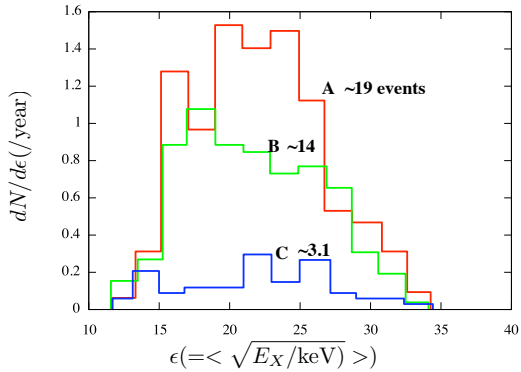} 
\caption{Comparison of expected signals by Model A, B and C}
\label{modelABC}
\end{center}
\end{figure}

Figure \ref{modelABC} compares the case of Model B and C with Model A (b12).
The number of expected events for Model B is $\sim 14$; it's smaller than model A ($\sim 19$)
but within the   detectable range.  The smaller number reflects the  difference of the spectrum
shape,
but it would be   difficult to derive the original spectrum shape.  Artificial model C gives
$\sim 3.2$ events in a year. This gives a reference of the minimum detectable flux;
we would be able to  say that a model is plausible  or  incompatible with the observation
only if  the flux is much higher than
$1.5\times 10^{-8}(E$/2.3TeV)$^{-3}$(/m$^2$s sr GeV) over 2 TeV. 

\subsubsection{Light/thermal shielding and pile-up effect}
In this trial calculation, we assume X-rays with more than 80 keV energy. 
However, we must prevent a pile-up effect in BGO due to  lower energy X-rays.
CFRP is supposed for light/thermal  shielding,  and if we make its thickness 
to be 500$\mu$m (0.11g/cm$^2$) carbon equivalent,  we obtain 
the X-ray transmission rate as shown in Table \ref{pileup}.

 \begin{table*}[htdp]
\caption{X-ray absorption by 500$\mu$m carbon}
\label{pileup}
\begin{center}
 \begin{tabular}{|c|c|c|c|c|c|c|c|c|c|c|}
 \hline
$E_x$ (keV)  &   \multicolumn{2}{|c|}{3}  &    \multicolumn{2}{|c|}{5}   &  \multicolumn{2}{|c|}{10}  & \multicolumn{2}{|c|}{60}   & \multicolumn{2}{|c|}{100}  \\
\hline
incident angle (deg) & 0  & 70  & 0  & 70  & 0  & 70 & 0  & 70 & 0 & 70\\
\hline
transmission rate(\%)  & $0.0012 $ & 0  &  25 & 1.5  & 75 & 43   & 98 & 95 &  98 & 95  \\
 \hline
\end{tabular}\\
 \end{center}
\end{table*}
CXB over 3 keV is expected in a 5 cm $\times$ 5 cm area at
$\sim $400 Hz,  but those photons reaching the  BGO are less than 100 Hz.
Low energy albedo  has much smaller flux than CXB,  as
in Fig. \ref{fig:cbxVSalbedo}.  Supposing the effective flux of albedo  over 80 keV
is 4 times  CXB,  then the rate of each BGO is  $\sim 70$ Hz.
Therefore,  X-rays entering  each BGO will not exceed 200 Hz
in normal conditions.   This means pile-up in BGO dose not matter
since the decay time of BGO light emission 
is $\sim 300$ ns.

\subsubsection{Trigger rate and dead time}
Trigger rate will be governed by  multi-Compton scattering.  
According to the simulation, the CXB contribution  with the condition
of any 3  BGO coincidence ($E_X>80$keV) is
$\sim$1Hz, and for  any 4 coincidence $\sim$0.045Hz.
The albedo will contribute a maximum of 100  times of this and 
will make   $\sim$100Hz for any 3 and $\sim 4.5$ Hz  for any 4 coincidence, 
respectively. 
Our target is $m\ge 5$, however, any 4 coincidence 
 trigger will be appropriate.

Since the decay time of BGO  light is $\sim$ 300 ns, we may assume
$\sim 1$ ms dead time.  Then, the trigger rate of $\sim$10 Hz presents  no
problem.

\section{Summary and concluding remarks}
\begin{itemize}
\item The purpose of this article is a study of  the feasibility of 
probing cosmic ray  primary electrons in the
$\sim$10 TeV region  by observing geo-synchrotron X-rays. 
According to some  plausible models,
such high energy electrons are expected to come from
supernova remnants such as Vela. 
\item  For this, we assumed a tentative detector design consisting of 
	      a number of BGO blocks as described in \ref{sec:detector}
\item  It is found that we cannot use events which contain the electron itself 
	falling on the detector; a high energy electron splashes   a number of  X/gamma-rays
	and thus it is difficult to identify aligned X-rays.  This reduces event the rate substantially.
\item   To realize a background free observation, 
      the number of triggered BGO's (with X-ray energy$>$ 80 keV)  must be $m\ge 5 $ and
	  they must  span more than 62 cm.
\item  It will be difficult to get a  spectrum shape for electrons.
 	However, owing to the exponential sharp cut-off of the  X-ray  energy
		spectrum for each electron energy, 
		  the  assumed detector  has no sensitivity to electrons
		below 2 TeV, and thus works like a threshold detector. 
\item  We can verify the plausibility  of  models  if the flux
		is much (4$\sim$5 times) higher than
		$1.5\times 10^{-8}(E$/2.3TeV)$^{-3}$(/m$^2$s sr GeV) over 2 TeV.
		The number of signal events is expected to be  15 to 20 in a year.

\item   It will be possible to increase the number of signal events by
  	 lowering  the threshold for observation down to
	$sim$50 keV without a big problem (chance coincidence is still
		negligible and   multi-Compton   scattering event rate is not affected).
		Then, the number of signal events will double or the detector
		could be made smaller.

\item  Lowering $m$ by 1 is very challenging  because the chance coincidence 
	probability increase substantially.  However, in our calculation we only considered
	$\Delta S$ for a line alignment constraint.  This is a loose condition rather  than a rigorous one.  
	Also we may note that $\epsilon $  by chance coincidence  distributes
	sharply at the lower  end of   the  true   signal distribution.
	  This is useful to separate some of  the true signal.  These factors are worth further  consideration.
\item The observation considered here would be more attractive if it could be combined
	with gamma-ray burst observations.

　
\end{itemize}

\section*{Acknowledgemnts}
This work was supported by JSF (Japanese Space Forum) as the 9th
public research for space utilization (2006$-$2008). We thank 
many concerned staff of JSF, especially T. Fujishima, for 
their assistance.  We also thank N.Hasebe  of Waseda Univ. 
for using an electronic  device.   Thanks are also due 
to J. P. Wefel of Louisiana State Univ.
for the carful reading of the manuscript and valuable
comments.

\appendix

 \section{Basic test of detector components}
\label{sec:basictest}
To get basic performance of assumed detector components, we performed the following
basic test experiments.  The parameters thus obtained were used in the present calculations.

\subsection{Materials}
\label{sec:material}
\begin{itemize}
\item  BGO (SICCAS)\footnote{%
	A LaBr3 (Sangoban) package
	 was partly  used as a reference, especially,  at low energies
	where BGO output  becomes  weak.  LaBr3's property:  density  5.29 g/cm$^3$, 
	radiation length 2.1, decay time 25 ns, 	 light yield 63 photons/keV or  200 times NaI.
}:

	  5cm$\times$5cm$\times$2cm t block $\times 8$. As reflector, 
	white teflon was used. Thin CFRP was used for light/thermal  shield. 
\item  PMT (Hamamatsu): 

	 R3318-HA(bialkali, 2''square) $\times 2$.\\
	R6231-100HA(super bialkali, 2''$\phi$) $\times 2$

	 To see possible individual differences,
		two PMT's were used for each type.
\item  X and gamma-ray source

	$^{241}$Am(60keV, 14 keV),  $^{57}$Co(122keV), $^{137}$Cs(662 keV), $^{60}$Co(1.17 MeV, 1.33MeV)

\end{itemize}

\subsection{Temperature dependence}
To avoid a temperature dependence in the test,  and for future applications, we examined the
temperature dependence of pulse height for the complete absorption peak of 662 keV gamma-rays.

At 15 to 30$^\circ$ C, temperature dependence of BGO  is within $\sim \pm 2$\%. 
At lower temperatures down
to -30$^\circ$C, $\sim -0.5$\%/deg  dependence is seen. Therefore, the temperature effect 
is negligible in our test conducted at room temperatures. LaBr3 has only $\sim \pm 2$\% changes
over the entire range.

\subsection{Energy resolution}
We exposed isotope sources  mentioned in \ref{sec:material} to the BGO's  to obtain pulse height
distributions.  Except for 60 keV X-rays from $^{241}$Am, data was taken for  all combinations
of  3 sources, 8 BGO blocks and 4 PMT's (total 96 cases).  

The maximum difference of energy resolution (in FWHM)
among 8 BGO's  is 4\% for 662 keV  and 7\% for 122 keV.   So we need to choose good BGO
for low energy X-ray observation in  actual  application. 
Two PMT's,  Bialkali (B) or  Super Bialkali (SB), show only a small difference (within 2\%).
SB  gives $\sim 2$ \% better resolution than B for 662 keV and $\sim 4$\% for 122 keV.
The SB  PMT recognized clearly two peaks at 1.17 MeV and 1.33 MeV gamma-rays from $^{60}$Co, but
the B PMT showed no clear separation.

This difference is reasonable considering the maximum yield wavelength (480 nm) of BGO,  
wave length dependence of the quantum  efficiency and  differences in light collection efficiency due to
the square and circular PMT areas.

\begin{figure}[htbp]
\begin{center}
\includegraphics[width=7cm]{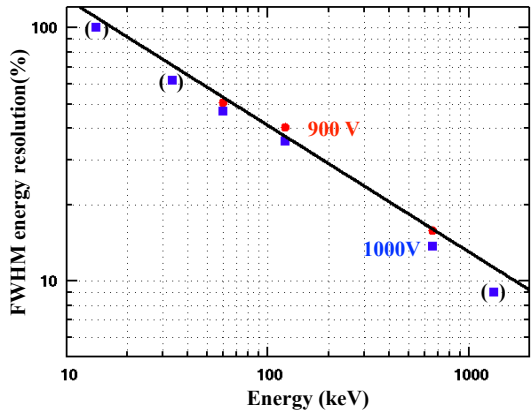} 
\caption{Energy dependence of energy resolution (\%  of FWHM). 
The results for SB PMT  voltage of  900 V and 1000 V are shown.
Accuracy of points with ()  is less than other points due to 
 difficult  background subtraction.    The point at $\sim 33$ keV is
 estimated from  a peak obtained in $^{57}$Co case 
 (Note: the K-shell energy  of Bi is 90.5 keV and $^{57}$Co has  $\sim$11\% 
 136 keV  line besides its 122 keV line. We  expect a line near $(123.5-90.5)$ keV.
 ). 
The curve shows the resolution,  $16\sqrt{662\ \rm{(keV)}/E})$ \%, used in the
simulation to put a random error on the absorbed energy in BGO.
}
\label{fig:Eresol}
\end{center}
\end{figure}

Figure \ref{fig:Eresol} shows  SB PMT measurement  results.    A 
difference of $\sim$3 \% does not matter  for our purpose and, in the simulation,
 we assumed a resolution
shown by the
curve in the figure to introduce a random error to the absorbed energy in
BGO.

\subsection{Dynamic range}
Dynamic range of the PMT was examined by applying voltages from 850 to 1000 V.  The pulse height 
is linear  as a function of  X-ray energy  from 14 keV to 1.25 MeV, except for  850 V case
for which linearity is lost in the 10 keV region (it is difficult to see the 14 keV peak).   
Voltage dependence of the  pulse height  is $\propto V^{5.6}$ for each energy. 
Since our target is 50 keV to  few MeV, this feature  will  make it easy to build electronics.

 \subsection{Time resolution}
 \begin{figure}[htbp]
\begin{center}
　\includegraphics[width=70mm]{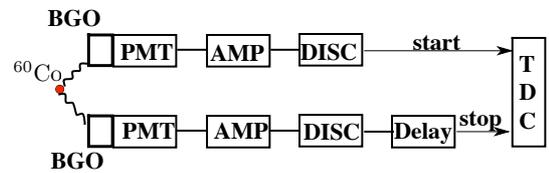}
  \caption{Setup for measuring time resolution of  the BGO+PMT system}
\label{fig:timingsetup}
\end{center}
\end{figure}
 To avoid chance coincidence, good time resolution is indispensable.   To see the time
 resolution,  we use a system as shown in Fig. \ref{fig:timingsetup} and measured time
 difference, $\Delta T$, from {\it start} to {\it stop}.
 Two gamma rays from
 $^{60}$Co (1.17 and 1.33 MeV) are emitted isotropically and simultaneously and will 
 enter BGO at the same  time.  In some cases, one gamma-ray which enters   a BGO block
 may  be Compton scattered and enters another BGO.  This timing is also the same within 
 our accuracy.

\begin{figure}[htbp]
\begin{center}
\includegraphics[width=70mm]{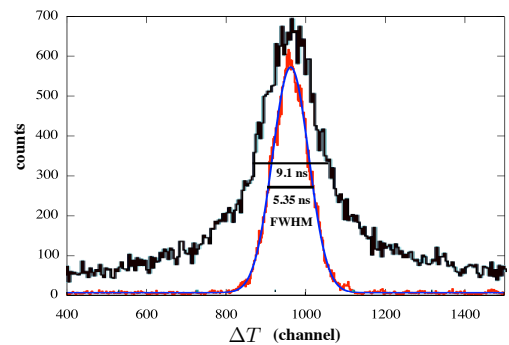}
  \caption{Time difference distribution observed in the setup shown in Fig. \ref{fig:timingsetup}.
  One channel is 50 ps. The constant fraction method gives 5.4 ns FWHM resolution, 
  while the leading edge method gives  9.1 ns, and the distribution has a  long tail deviating from Gaussian.
}
\label{fig:timing}
\end{center}
\end{figure}

To get good resolution, we  use a constant fraction discriminator (ORTEC 935). 
We also use a leading edge discriminator  to confirm the merit  of using the
constant fraction method as seen in Fig. \ref{fig:timing}

The pulse height from 1.2 MeV absorption is $\sim$ 100 mV, and we 
use a discrimination level  of 50 mV. For timing, 50\% fraction is used.
The data include Compton
events which have  smaller energy than 1.2 MeV; thus the result indicates
similar  good  time resolution  even for smaller energy deposit.
To see this, we use the discrimination level of 25 mV and 10  mV to find
 less than 1 ns degradation of the time resolution.
This suggests 10 ns resolution is quite safe for several tens
keV region.

\end{document}